\documentclass[10pt,twocolumn,letterpaper]{article}

\usepackage{iccv}
\usepackage{times}
\usepackage{epsfig}
\usepackage{graphicx}
\usepackage{amsmath}
\usepackage{amssymb}
\usepackage[accsupp]{axessibility}

\usepackage[pagebackref=true,breaklinks=true,letterpaper=true,colorlinks,bookmarks=false]{hyperref}

\iccvfinalcopy 

\def\iccvPaperID{34} 

\ificcvfinal\fi

\begin{document}

\title{Simple and Efficient Unpaired Real-world Super-Resolution using Image Statistics}

\author{Kwangjin Yoon\\
SI Analytics\\
70 Yuseong-daero, Yuseong-gu, Daejeon 34047, Republic of Korea\\
{\tt\small yoon28@si-analytics.ai}
}

\maketitle
\ificcvfinal\thispagestyle{empty}\fi

\begin{abstract}
Learning super-resolution (SR) network without the paired low resolution (LR) and high resolution (HR) image is difficult because direct supervision through the corresponding HR counterpart is unavailable. Recently, many real-world SR researches take advantage of the unpaired image-to-image translation technique. That is, they used two or more generative adversarial networks (GANs), each of which translates images from one domain to another domain, \eg, translates images from the HR domain to the LR domain. However, it is not easy to stably learn such a translation with GANs using unpaired data. In this study, we present a simple and efficient method of training of real-world SR network. To stably train the network, we use statistics of an image patch, such as means and variances. Our real-world SR framework consists of two GANs, one for translating HR images to LR images (degradation task) and the other for translating LR to HR (SR task). We argue that the unpaired image translation using GANs can be learned efficiently with our proposed data sampling strategy, namely, variance matching. We test our method on the NTIRE 2020 real-world SR dataset. Our method outperforms the current state-of-the-art method in terms of the SSIM metric as well as produces comparable results on the LPIPS metric.
\end{abstract}

\section{Introduction}	
\label{sec:intro}	

\begin{figure}[t]	
    \centering	
    \includegraphics[width=1.\linewidth]{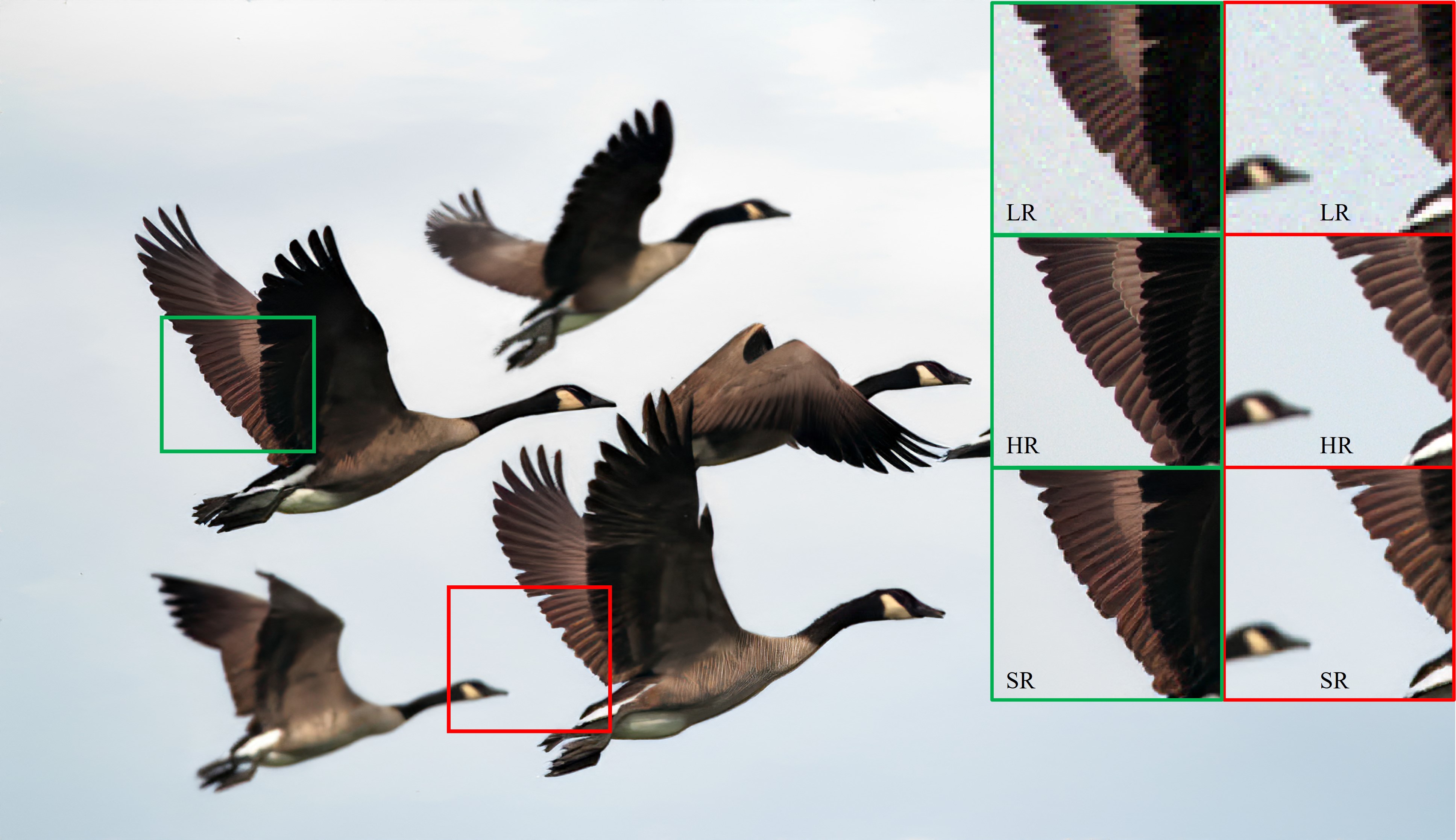}	
    \caption{Result of super-resolved image ($\times 4$) by the proposed method.}	
    \label{fig:main}	
\end{figure}

Single image Super-Resolution (SR) is the task of increasing the resolution of a given image as well as sharpening its content by predicting the high-frequency component and the missing information. Many types of research \cite{dong2015image, zhang2018image, ledig2017photo, wang2018esrgan} were conducted on paired low-resolution (LR) and high-resolution (HR) images to train the neural network in a fully supervised manner. They artificially produced paired data by downscaling HR images with known kernels (\eg bicubic) to generate corresponding LR images. Since there is a domain gap between downscaled images and real LR images, a network trained on downscaled images struggles to generalize to natural images \cite{lugmayr2020ntire}. On the other hand, some recent approaches tried to solve real-world SR problems by building real-world datasets where paired LR-HR images on the same scene are captured by adjusting the focal length of a digital camera \cite{cai2019toward, zhang2019zoom}. However, collecting such images is very expensive and time-consuming and requires extra conditions, namely non-moving objects in a scene.

Another stream of researches begins to focus on finding an actual degradation function, \ie{ translating image domain from HR to real LR} \cite{bulat2018learn, lugmayr2019unsupervised, yuan2018unsupervised}. In other words, many real-world SR researches have recently emerged that take advantage of the unpaired image-to-image translation framework \cite{zhu2017unpaired}. They used two or more generative adversarial networks (GANs), each of which translates images from one domain to another domain, \eg, translating a given HR image to an LR image that is not distinguishable from a real LR image. However, it is difficult to learn such translations using GANs, particularly with unpaired data.

This study proposes a simple and efficient training method for real-world SR that takes unpaired data as the input. We show that our method is able to improve the SR network by utilizing the statistics of an image patch. Specifically, the method restricts the amount of difference of the variance between LR and HR patches. By doing so, we can match the content density of those patches; hence a network can learn from unpaired patches that have a similar level of content.

We present the related work in section \ref{sec:related}. Then, our method is described in detail in section \ref{sec:method}. In section \ref{sec:exp}, we test our method on the NTIRE 2020 real-world SR dataset by performing $\times4$ SR task for each dimension. Our method outperforms the current state-of-the-art method in terms of the SSIM metric as well as produces comparable results on the LPIPS \cite{zhang2018unreasonable} metric. Finally, we concluded the paper in section \ref{sec:conclusion}.	

\section{Related work}	
\label{sec:related}

\begin{figure*}[ht]	
    \centering	
    \includegraphics[width=1\linewidth]{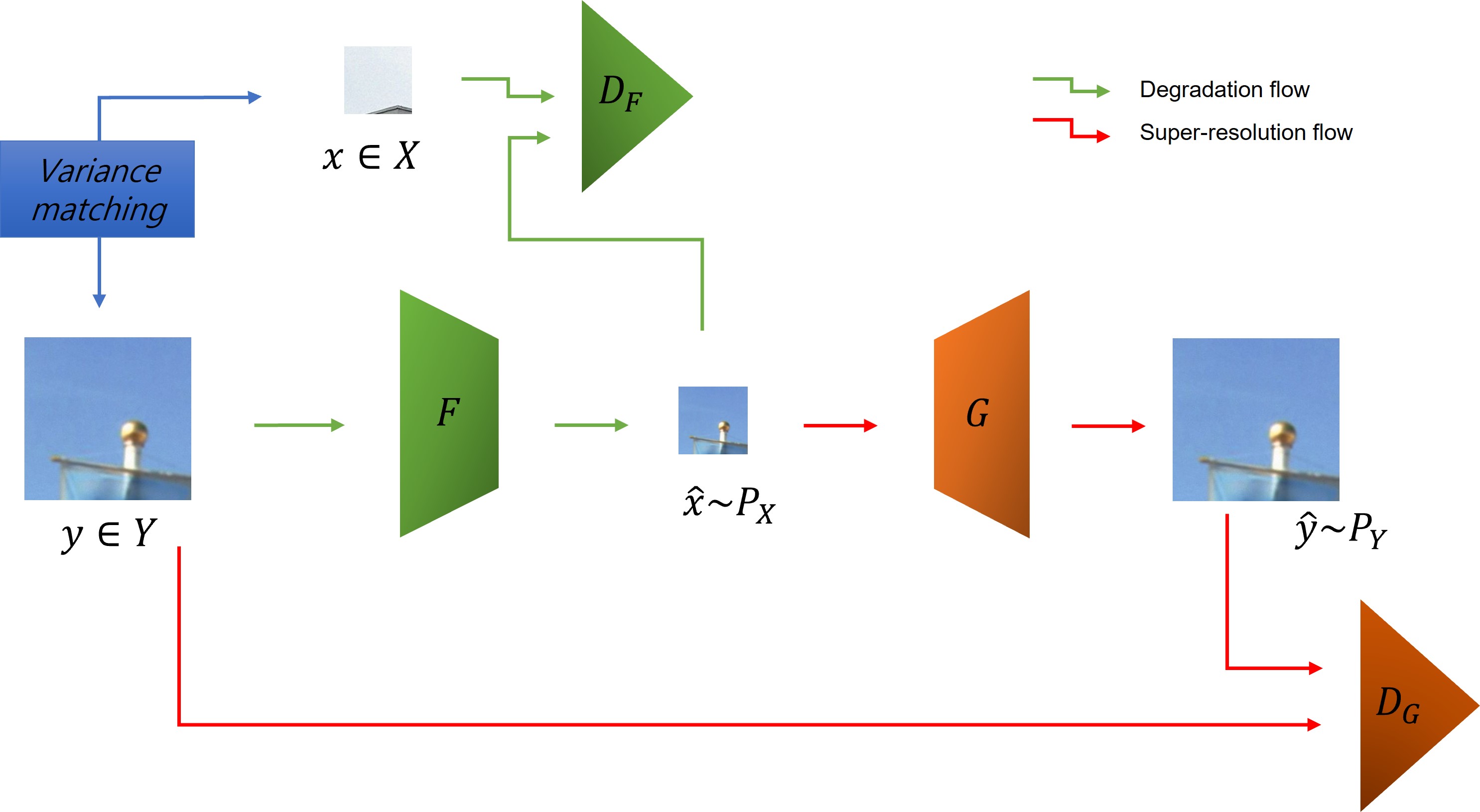}	
    \caption{Framework and data flow of our method. $F$ degrades HR images into the real LR images, and $G$ super-resolves the LR images. \textit{Variance matching} helps the training procedure of two GANs by constraining the variance difference between LR and HR patches so that they have a similar level of content. Detailed explanations are in section \ref{sec:sampling}. For simplicity, the loss components are omitted.}
    \label{fig:arch}	
\end{figure*}	

Recent super-resolution researches achieve strong performance on bicubic downsampled low-resolution images thanks to the convolutional neural network (CNN) \cite{zhang2018image, ledig2017photo, wang2018esrgan}. Zhang \etal \cite{zhang2018image} proposed very deep residual channel attention network (RCAN), which uses a residual in residual structure to form a very deep network. Ledig \etal \cite{ledig2017photo} used the residual connection mechanism to construct a deeper network with perceptual losses \cite{johnson2016perceptual} and generative adversarial network (GAN) \cite{goodfellow2014generative} which is called SRGAN. Wang \etal \cite{wang2018esrgan} introduced the enhanced SRGAN (ESRGAN), which improved the SRGAN by enhancing its architecture and perceptual loss. However, these SR models produce a poor result when a real-world LR image is inputted because they were trained on the paired data whose LR images were simply generated by downsampling HR counterparts with bicubic kernel. To address this issue, some researches \cite{cai2019toward, zhang2019zoom} collected the LR-HR pairs directly using particular camera instruments and introduced the real-world paired datasets. 	

Collecting the paired data requires an immense cost and extra conditions, such as a scene with non-moving objects. To overcome the problem, several real-world SR researches \cite{bulat2018learn, lugmayr2019unsupervised, yuan2018unsupervised, fritsche2019frequency} are proposed which use GANs to learn the conditional distribution of LR domain given an HR image. Bulat \etal \cite{bulat2018learn} proposed a two-stage process which firstly trains a high-to-low GAN with unpaired HR and LR images, then the output of the network is inputted to low-to-high GAN, which is trained for super-resolution using paired LR and HR images. Lugmayr \etal \cite{lugmayr2019unsupervised} also used CycleGAN to translate HR images to LR images and consequently constructed the paired LR and HR dataset that is used for training the SR network. Yuan \etal \cite{yuan2018unsupervised} proposed a cycle-in-cycle network to simultaneously learn a degradation (high-to-low) network and SR network (low-to-high). In \cite{fritsche2019frequency}, Fritsche \etal proposed to separate the low and high image frequencies and treat them differently during training. The high-frequency information of an image is adversarially trained with a discriminator, and the low-frequency information is learned with the $L1$ criterion. These methods use two or more GANs, each of which translates a set of images of one domain to the other domain. However, training GAN with unpaired data is prone to unstable. Our method can learn such image translation efficiently with simple image statistics.	

In \cite{ji2020real}, Ji \etal proposed a novel degradation framework for real-world images by estimating various blur kernels as well as real noise distribution. In particular, they extracted noise patches from the real-world LR images whose variance is lower than a threshold. Then the extracted noise patch is added to downsampled HR image to imitate the real-world LR image. The assumption of their method is that the variance alone is enough to decouple noise and content. However, applying their method of generating LR images to other datasets needs far more effort than the learning-based methods since they engineered an image processing technique for a specific dataset. Furthermore, the existence of the threshold parameter needs sufficient prior knowledge about noise which is empirically set.	

\section{Method}	
\label{sec:method}	

In this section, we introduce the proposed method. A brief description of our unpaired SR framework and data flow is depicted in Figure \ref{fig:arch}. As shown in the figure, our network consists of two GANs. The detailed description of the architecture is presented in section \ref{sec:net}. 

Let $x$ be an image in the LR domain $X$ and $y$ be an image in the HR domain $Y$. Then, we learn a mapping $F:Y \rightarrow X$ using a GAN such that the output $\hat{x} = F(y), y \in Y$, is indistinguishable from images $x \in X$, \ie{ $\hat{x} \sim P_X$, where $P_X$ is the distribution of $X$}. In addition, we also learn another mapping $G:X\rightarrow Y$ with a GAN such that the output $\hat{y} = G(x), x \in X$, belongs to $Y$, \ie{ $\hat{y} \sim P_Y$}. Therefore, $F$ degrades HR images into the real LR images, and $G$ super-resolves the LR images. These two mappings are end-to-end trained with unpaired data. Since each of the two mappings is trained with the GAN framework, there are two adversarial discriminators, $D_F$ and $D_G$ for each generator, respectively. 

In order to learn such mappings better, we also propose a simple and efficient training method which is described in section \ref{sec:sampling}.


\subsection{Network Architectures}
\label{sec:net}

In this section, we describe the architecture of our network which is depicted in Fig \ref{fig:arch}. Our network architectures are adopted from previous works \cite{wang2018esrgan, zhu2017unpaired}. We used ESRGAN architecture \cite{wang2018esrgan} for generators. The network $G$ consists of RRDBs \cite{wang2018esrgan} and two $\times 2$ upsampling layers, resulting in $\times 4$ SR network. Generator $F$, which translates images from domain $Y$ to $X$ (degradation task), also adopts ESRGAN with modifications replacing upsampling layers with average pooling layers. For the discriminator networks, $D_G$ and $D_F$, we use PatchGANs \cite{zhu2017unpaired} which aim to classify whether $70 \times 70$ overlapping image patches are real or fake.	

\subsection{Losses}
\label{sec:loss}

Our objective for both generators $F$ and $G$ contains the adversarial loss \cite{goodfellow2014generative}, the perceptual loss \cite{johnson2016perceptual} and the feature matching loss \cite{salimans2016improved}. The cycle consistency loss \cite{zhu2017unpaired} is used for training $G$ while generator $F$ applies a content loss which is $L_1$ distance in the pixel domain. Therefore, the image translation cycle with our generators is only learned with the forward cycle consistency \cite{zhu2017unpaired}. The objective of discriminators, $D_G$ and $D_F$, is computed by LSGAN \cite{mao2016least} discriminating real data from generated data.	

Specifically, for the generator $G$, we optimize a loss:
\begin{equation}
    L_G = \lambda_G^{adv}L_G^{adv} + \lambda_G^{cyc}L_G^{cyc} + \lambda_G^{per}L_G^{per} + \lambda_G^{fea}L_G^{fea},
\end{equation}
where $L_{G}^{adv}$ is the adversarial loss \cite{mao2016least}. $L_{G}^{cyc}$ is the cycle consistency loss \cite{zhu2017unpaired} which is a $L_1$ distance between $y$ and $G(F(y))$. We used the improved perceptual loss ($L_G^{per}$) inspired by \cite{wang2018esrgan}. In addition, $L_G^{fea}$ is the feature matching loss \cite{salimans2016improved} that is a $L_2$ distance between feature vectors $D_G^f(y)$ and $D_G^f(G(\hat{x}))$, where $D_G^f$ is an $f$-th intermediate layer of the discriminator $D_G$. Finally, $\lambda_G^{adv}$, $\lambda_G^{cyc}$, $\lambda_G^{per}$ and $\lambda_G^{fea}$ are weight constants for corresponding losses.

For the generator $F$, the following loss is optimized:
\begin{equation}
    L_F = \lambda_F^{adv}L_F^{adv} + \lambda_F^{con}L_F^{con} + \lambda_F^{per}L_F^{per} + \lambda_F^{fea}L_F^{fea}.
\end{equation}
The adversarial loss ($L_F^{adv}$), perceptual loss ($L_F^{per}$) and feature matching loss ($L_F^{fea}$) are applied in a similar way to $G$. Furthermore, instead of the cycle consistency loss used in $G$, $L_F^{con}$ is applied that is $||B(y) - F(y)||_1$. $B(\cdot)$ is a bicubic downsampling operation. Similarly, $\lambda_F$ with the particular superscript is a weight constant for the corresponding loss.

\subsection{Sampling with Image Statistics}	
\label{sec:sampling}

We observed that learning the image translation task for SR with the unpaired dataset is difficult. We frequently observe that the procedure ends up in a trivial solution, \ie produces an output that is merely a simple linear interpolation of a given input or fails to converge. We conjecture that the phenomenon is caused by random sampling from two independent variables, $X$, and $Y$. The unpaired input $x$ and $y$ are independent of each other, and hence each input contains completely unrelated contents and has varying degrees of image statics between them. To solve the problem, we propose to use image statistics when sampling LR and HR image patches. Specifically, we used the variance of the image patch. Let $\sigma_{x}^2$ and $\sigma_{y}^2$ be the variance of LR patch $x$ and HR patch $y$, respectively. After randomly sampling $x$, an HR patch $y$ is retrieved that satisfying the following condition:	
\begin{equation}	
    | \sigma_x^2 - \sigma_y^2 | < \sigma_T^2	
    \label{eq:1}	
\end{equation}	
where, $\sigma_T^2$ is a threshold value that controls the difference of variance between $x$ and $y$. We call this \textit{variance matching}. 

\begin{figure}[t]	
    \centering	
    \includegraphics[width=1\linewidth]{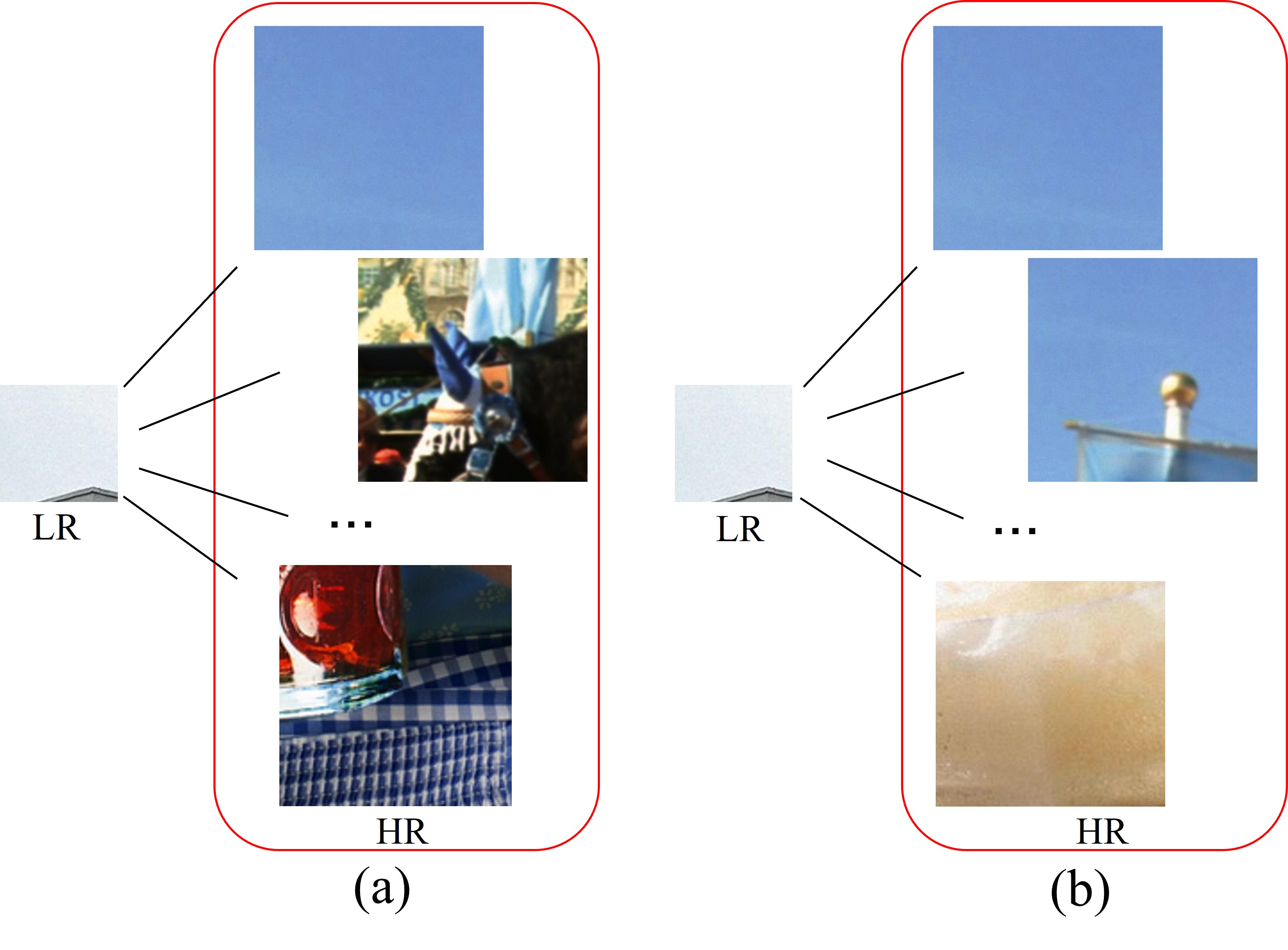}	
    \caption{Exemplar diagram of our sampling method. An example LR patch has a low variance, where a large area is filled with the sky, and a roof of a building takes a tiny area at the bottom. Possible HR counterparts for the LR patch are shown in red rectangles. (a) independent sampling: There is no restriction for sampling an HR patch. (b) our method: an HR patch is sampled, of which variance is determined by the given LR patch.}
    \label{fig:sample}	
\end{figure}

Here, we used variance because the variance of an image patch is related to the content itself \eg a patch with large variance has rich content \cite{ji2020real}. By restricting the amount of difference of the variance between LR and HR patches, we can match the content density of those patches. Hence a network can learn from unpaired patches that have a similar level of content (Figure \ref{fig:sample}).

The variance matching at first helps the training of the degradation networks, $F$ and $D_F$, by constraining the variance difference between HR and LR patches which are inputted $F$ and $D_F$, respectively. So, they are trained with patches that have a similar level of content. Sequentially, the super-resolution networks ($G$, $D_G$) are also properly trained with $\hat{x}$ that is indistinguishable from $X$ (Figure \ref{fig:arch}). Again, our networks are trained in an end-to-end manner.


\section{Experiments}	
\label{sec:exp}	

We test our method on the NTIRE 2020 real-world SR dataset \cite{lugmayr2020ntire}. LR part of the training set is constructed by applying the degradation operation to the $2650$ images of the Flickr2K \cite{wang2018esrgan} dataset. The degradation operation is undisclosed to the public. HR part of the training set is composed of the original $800$ training images from DIV2K \cite{timofte2017ntire} dataset. The validation set of the dataset consists of paired LR-HR images from the validation split of DIV2K, where the LR images are obtained by first downscaling the HR counterparts followed by the degradation. We did not use the test set of the dataset for the quantitative analysis since the ground truth of the set can not be accessed.	

\subsection{Implementation Details}	

The learning rates of $G$, $F$, $D_G$ and $D_F$ are initially set to 0.0001 and are halved at 100, 200, 400 and 700 epoch. For $G$, weight constants $\lambda_G^{adv}$, $\lambda_G^{cyc}$, $\lambda_G^{per}$ and $\lambda_G^{fea}$ are set to 0.3, 0.2, 0.5 and 20.0, respectively. For $F$ $\lambda_F^{adv}$, $\lambda_F^{con}$, $\lambda_F^{per}$ and $\lambda_F^{fea}$ are set to 0.3, 0.5, 0.2 and 20.0, respectively. We used $64$ for the variance threshold $\sigma_T^2$. The size of input patch of $G$ is $32^2$ and that of $F$ is $128^2$.	

\subsection{Sampling strategy}	

If we naively collect an HR patch after an LR patch is chosen, then it takes uneven times to sample an HR patch. This is because we have to search for an HR patch that satisfies the equation \ref{eq:1}. We simply circumvent this issue and make the sampling time rather consistent. We first randomly collect a bunch of LR patches, namely $N_{LR}$ patches, from an LR image. Similarly, $N_{HR}$ HR patches are taken from an HR image. Then, the variance of each patch is calculated, and we finally take pairs that satisfy the equation \ref{eq:1} among $N_{LR} \times N_{HR}$ pairs. The law of large numbers inspires the idea of this simple sampling strategy. We used $30$ for both $N_{LR}$ and $N_{HR}$, which is larger than the batch size.	

\subsection{Quantitative Results}	

We compared our method with ESRGAN \cite{wang2018esrgan}, ZSSR \cite{shocher2018zero}, K-ZSSR \cite{bell2019blind}, CinC \cite{yuan2018unsupervised} and Ji \etal \cite{ji2020real} in Table \ref{tab:1}. Ours$^-$ refers to the proposed method without the variance matching while Ours$^+$ refers to the full implementation. Ji \etal took the first place in NTIRE 2020 Real-world SR challenge. Ours$^+$ achieved the best SSIM score ($\uparrow$) which is $0.01$ higher than Ji \etal, and the second best on the LPIPS \cite{zhang2018unreasonable} metric ($\downarrow$) which is $0.055$ higher than the first ranker (Ji \etal). 

It is noteworthy that, with the variance matching, the performance is considerably improved as Ours$^+$ outperforms Ours$^-$ at all three metrics in Table \ref{tab:1}. This indicates that if cycle-GAN-like real-world SR methods \cite{bulat2018learn, lugmayr2019unsupervised, yuan2018unsupervised} are equipped with the variance matching, their performance will be further improved. In addition, applying the variance matching to existing methods is very simple and effortless since it is only involved in the data sampling.

\begin{table}[t]	
\centering	
\begin{tabular}{c c c c} 	
 \hline	
 Method & PSNR $\uparrow$ & SSIM $\uparrow$ & LPIPS $\downarrow$ \\	
 \hline\hline	
 ESRGAN \cite{wang2018esrgan} & 19.06 & 0.2423 & 0.7552 \\ 	
 ZSSR \cite{shocher2018zero} & \textbf{25.13} & 0.6268 & 0.6160 \\	
 K-ZSSR \cite{bell2019blind} & 18.46 & 0.3826 & 0.7307 \\
 CinC \cite{yuan2018unsupervised} & 24.05 & 0.6583 & 0.4593 \\
 Ji \etal \cite{ji2020real} & 24.82 & 0.6619 & \textbf{0.2270} \\	
 \hline	
 Ours$^-$ & 23.01 & 0.6389 & 0.3183 \\	
 Ours$^+$ & 24.30 & \textbf{0.6731} & 0.2824 \\ 	
 \hline	
\end{tabular}	
\caption{Quantitative results on the validation set compared with ESRGAN, ZSSR, K-ZSSR, CinC and Ji \etal. Ours$^-$ refers to the proposed method without the variance matching while Ours$^+$ refers to the full implementation.}	
\label{tab:1}	
\end{table}	

\subsection{Ablation study}	

\subsubsection{$\sigma_T^2$ of the variance matching}

\begin{table}[ht]	
\centering	
\begin{tabular}{c | c | c} 	
 \hline	
 $\sigma_T^2$ & SSIM $\uparrow$ & LPIPS $\downarrow$ \\	
 \hline \hline	
 N/A   & 0.6389 & 0.3183 \\	
 $576$ & 0.6417 & 0.3021 \\	
 $256$ & 0.6545 & 0.3115 \\	
 $100$ & 0.6632 & 0.2989 \\ 	
 $64$  & \textbf{0.6731} & 0.2824 \\ 	
 $36$  & 0.6698 & \textbf{0.2822} \\	
 \hline	
\end{tabular}	
\caption{SSIM and LPIPS against the threshold $\sigma_T^2$.}	
\label{tab:2}	
\end{table}	

Here, we conducted an ablation study of the threshold value $\sigma_T^2$. We reported the SSIM and LPIPS scores while we adjusted $\sigma_T^2$ in table \ref{tab:2}, where N/A means that no variance matching was applied (Ours$^-$). As shown in the table \ref{tab:2}, the performance of SSIM was increased if we applied the variance matching. $\sigma_T^2 = 64$ produced the best result. It is until $\sigma_T^2=64$, the performance of SSIM increases as $\sigma_T^2$ decreases. However, if we further decrease $\sigma_T^2$, the performance of SSIM is be dropped, \ie, SSIM is dropped by $0.003$ at $\sigma_T^2 = 36$ compared to $\sigma_T^2 = 64$. 

For the LPIPS \cite{zhang2018unreasonable} metric, $\sigma^2_T = 36$ shows the most improved result with $0.2822$. However, it is worth noting that the performance gap between the best ($\sigma^2_T = 36$) and second-best ($\sigma^2_T = 64$) is only $0.0002$, while the best one requires more time for sampling. In addition, the LPIPS has relatively fluctuated if compared with the SSIM.

\subsubsection{Using mean as an image statistic}

We also considered the mean of a patch for the image statistics. Together with the constraint in equation \ref{eq:1}, the mean of an image patch is also used:
\begin{equation}	
    | \sigma_x^2 - \sigma_y^2 | < \sigma_T^2 \wedge | \mu_x - \mu_y | < \mu_T,
    \label{eq:2}	
\end{equation}
where $\mu_x$ and $\mu_y$ are the mean of LR patch $x$ and HR patch $y$, respectively. $\mu_T$ is a threshold value.

However, we found that using constraint \ref{eq:2} barely improved the performance. The results are shown in Table \ref{tab:2}, where $\sigma_T^2=64$ is fixed and $\mu_T$ is decreased at interval of $100$\footnote{The pixel values are ranged in $[0, 255]$. Thus, $\mu_T=255$ does not differ from N/A.}. It, on occasion, worsen the performance, \ie{ $\mu_T=155$}. At $\mu_T=55$, SSIM is very slightly increased by $0.0005$. It is worth noting that the sampling time of $\mu_T=55$ is much slower than not using it. It takes about $\times2.5$ more time for an iteration due to the added constraint.

In addition, using the mean only for the image statistic was also examined. However, we did not observe any performance improvement. This is because that using only mean does not guarantee that the matched patches have a similar level of content. Therefore, we mainly use variance matching (Eq. \ref{eq:1}).

\begin{table}[ht]	
\centering	
\begin{tabular}{c | c} 	
 \hline	
 $\mu_T$ & SSIM $\uparrow$ \\	
 \hline \hline	
 N/A   & 0.6731 \\	
 $155$ & 0.6729 \\	
 $55$ & \textbf{0.6736} \\
 \hline	
\end{tabular}	
\caption{SSIM score against the threshold $\mu_T$. $\sigma_T^2$ is fixed to $64$.}	
\label{tab:3}	
\end{table}	

\subsection{Qualitative Results}	

\begin{figure*}[ht]	
    \centering	
    \includegraphics[width=1\linewidth]{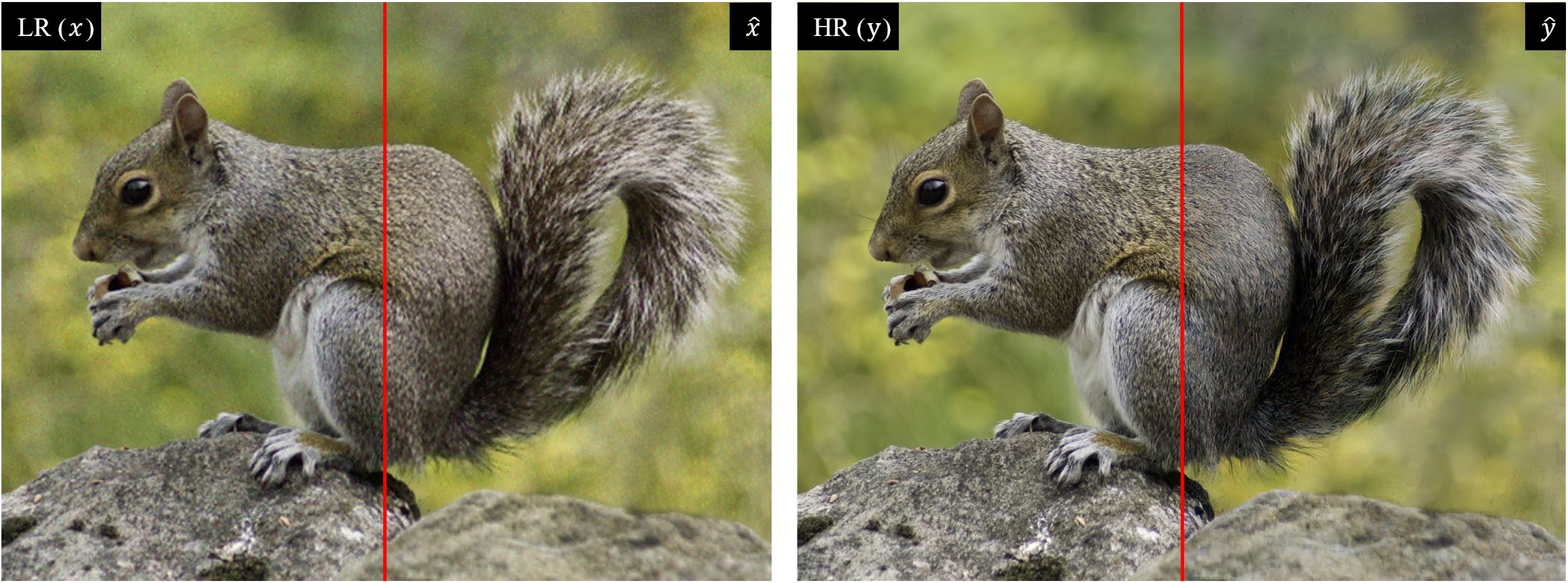}	
    \caption{Qualitative comparisons between real LR image $x$ and the degraded image $\hat{x}=F(y)$ (right), and between HR image $y$ and the super-resolved image $\hat{y}=G(x)$ (left).}	
    \label{fig:qual2}	
\end{figure*}	

In this section, we show the qualitative results. In Figure \ref{fig:qual2}, the results of image translation using our method are shown. As shown in the left side of Figure \ref{fig:qual2}, we compare the half of real LR image $x$ and the half of generated LR image $\hat{x}$ side-by-side. $\hat{x}$, the degraded result of $y$ by $F(y)$, successfully mimics the real noises. Namely, it is hard to distinguish $\hat{x}$ from $x$. On the right side of Figure \ref{fig:qual2}, we also presented the side-by-side comparison between HR image $y$ and super-resolved image $\hat{y}$. $\hat{y}$ is the SR result generated from a real-world image $x$, \ie $G(x)$. 

\begin{figure*}[ht]	
    \centering	
    \includegraphics[width=1\linewidth]{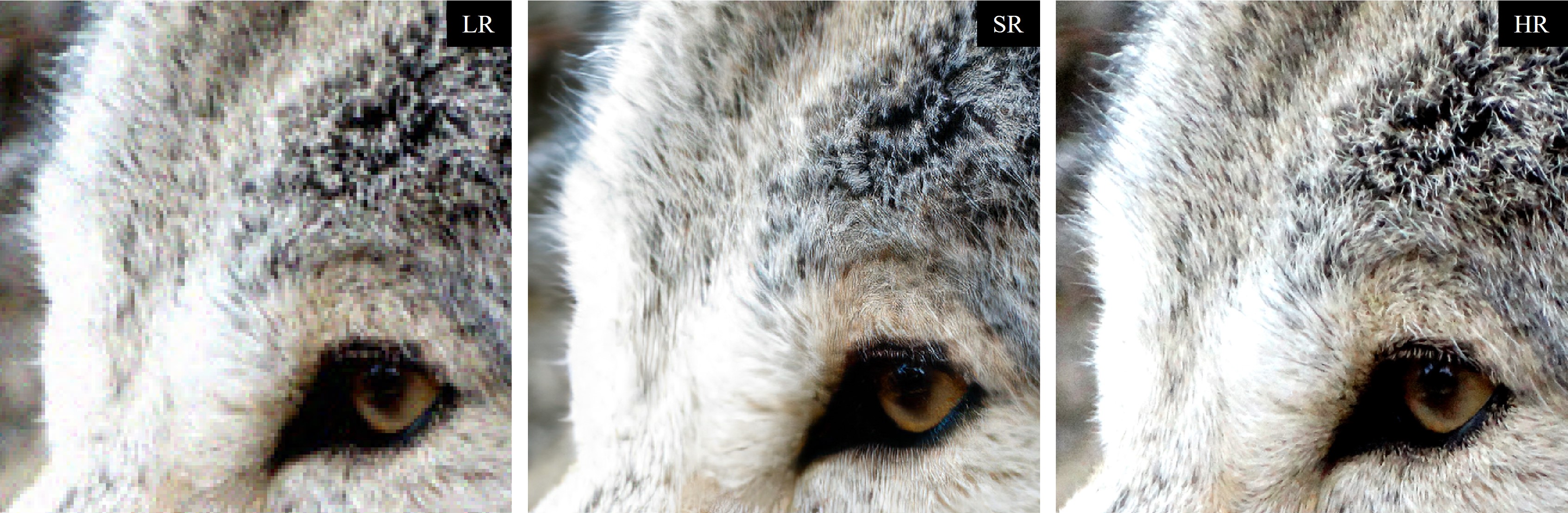}
    \hfill \break
    \includegraphics[width=1\linewidth]{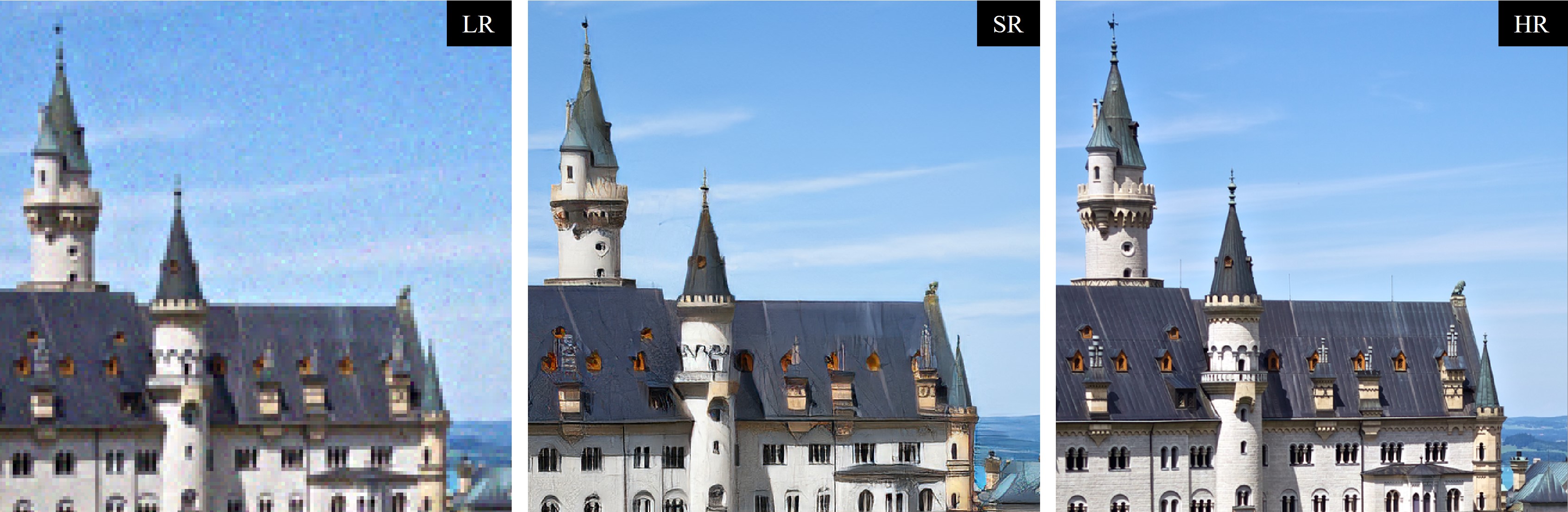}	
    \caption{SR results of our method (middle column). LR (left column) and HR (right column) are also shown.}	
    \label{fig:qual}	
\end{figure*}

\begin{figure*}[ht]	
    \centering	
    \includegraphics[width=1\linewidth]{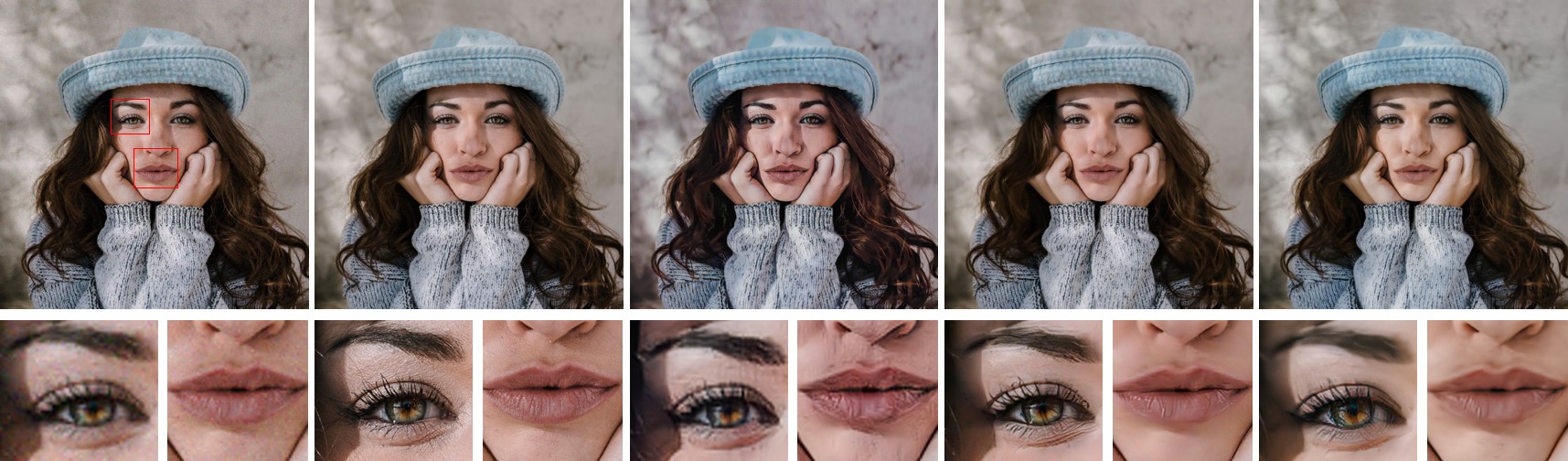}
    \hfill \break
    \includegraphics[width=1\linewidth]{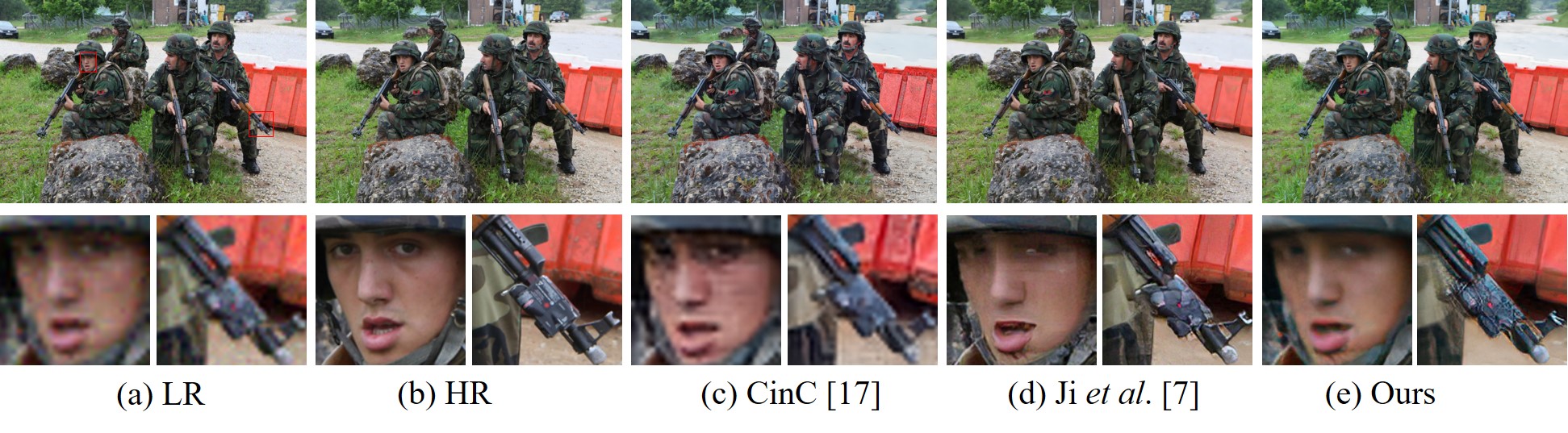}	
    \caption{Visualization comparison among CinC, Ji \etal, and our method with the variance matching. The red areas in LR images are zoomed in for the purpose of easier comparison.}	
    \label{fig:qual_vs}	
\end{figure*}

In Figure \ref{fig:qual}, SR results of our method are depicted in the middle column. Our method produces natural texture, \eg animal fur (the first row), as well as removes noise artifacts, \eg sky (the second row). Lastly, visual comparisons among CinC \cite{yuan2018unsupervised}, Ji \etal \cite{ji2020real} and ours are presented in Figure \ref{fig:qual_vs}.

\section{Conclusion}	
\label{sec:conclusion}

In this study, we proposed a simple and efficient unpaired real-world SR method that utilized the variance of image patches so that it can match the content density between LR and HR pairs. An HR patch $y$ is found after LR patch $x$ is randomly sampled while it satisfies the variance matching condition. With this strategy, we stably train an unpaired SR network, which can produce real-world SR results with better performance. We test our method on the NTIRE 2020 real-world SR dataset. Our method outperforms the current state-of-the-art method in terms of the SSIM metric.

{\small
\bibliographystyle{ieee_fullname}
\bibliography{main}
}
\end{document}